\documentclass[10pt]{article}
\usepackage{graphicx}
\usepackage{amsmath}
\usepackage{amssymb}
\usepackage{caption2}
\begin{document}
\bibliographystyle{prsty}
\begin{center}
{\large {\bf \sc{Radiative transitions  among the vector and scalar heavy quarkonium states with covariant light-front quark model }}} \\[2mm]
Zhi-Gang Wang  \footnote{E-mail:zgwang@aliyun.com. } \\
  Department of Physics, North China Electric Power University, Baoding 071003, P. R.
  China
\end{center}

\begin{abstract}
In  this article, we study the radiative transitions among the vector and scalar heavy  quarkonium states with the covariant light-front quark model.
In calculations, we observe that the radiative decay widths are sensitive to the constituent quark masses and the shape parameters of the  wave-functions,
 and reproduce the experimental data  with suitable parameters.
\end{abstract}


\section{Introduction}

Recently, the BESIII collaboration observed  the first evidence for  direct two-photon transition $\psi^{\prime}\to  J/\psi \gamma\gamma$ with the branching fraction     $(3.3\pm0.6 {}^{+0.8}_{-1.1} ) \times10^{-4}$ in a sample of 106 million $\psi^{\prime}$ decays collected by the BESIII detector \cite{BES1204}.
The branching fractions of the double $E1$ transitions $\psi^{\prime} \to ( J/\psi\gamma )_{\chi_{cj}}\gamma$ through  the intermediate states $\chi_{cj}$ ($j=0,1,2$) are also reported \cite{BES1204}, while previous experimental  data indicates that the double radiative decays $\psi^{\prime} \to  J/\psi\gamma\gamma$ take place through the decay cascades $\psi^{\prime} \to \chi_{cj}\gamma\to  J/\psi\gamma\gamma$  with  tiny non-resonance contributions \cite{CLEO-05,CLEO-08}.
In Ref.\cite{He1012}, He et al study the discrete contributions to decays $\psi^{\prime}\to J/\psi\gamma\gamma$  due
to the $E1$ transitions using the heavy quarkonium effective Lagrangian \cite{GattoPRT}. No theoretical work on the non-resonance's contributions exists.
On the other hand, we expect that there are non-resonance's contributions to the doubly radiative decays among the bottomnium states $\Upsilon(nS)$,
the doubly radiative transition $\Upsilon^{\prime\prime} \to \Upsilon^{\prime }\gamma\gamma$ has been observed \cite{PDG}.

The radiative transitions among the heavy quarkonium states are usually calculated model-dependently by the nonrelativistic potential quark models with considerable relativistic corrections \cite{RDecay-potential}, or calculated model-independently by the lattice QCD \cite{RDecay-Latt} and effective field theory \cite{RDecay-EFT}, one can consult the recent review \cite{RD-Review} for more references.  In general,  we expect to study both the resonance's and non-resonance's contributions in the doubly radiative transitions with the covariant light-front quark model (CLFQM), where the  wave-functions are expressed in terms of the internal variables of the quark and gluon, and maintain Lorentz covariance.
In Refs.\cite{Jaus1999,Jaus2003}, Jaus  introduces  the
CLFQM and takes into account the zero-mode contributions
systematically to preserve   covariance and remove   dependence of physical
quantities on the  light-front direction.  The  CLFQM has been  successfully  applied to calculate
 the $S$-wave and $P$-wave meson's decay constants and form-factors \cite{Cheng2003,Cheng2004,LFQM}.
 In Refs.\cite{Wang1002,Ke1002}, the authors study the $M1$ transitions $\Upsilon(nS)\to  \eta_b(n'S)\,\gamma$ in the CLFQM, and observe that  the   $M1$ transitions are sensitive to the heavy quark masses and shape parameters of the light-front wave-functions (LFWF), the existing experimental data cannot be reproduced consistently with suitable heavy quark masses and shape parameters \cite{Wang1002}. It is interesting to study whether or not the CLFQM can be successfully applied to calculate the $E1$ transitions among the heavy quarkonium states.

 In the nonrelativistic limit, the   decay widths of the $M1$ and $E1$ transitions are proportional to the squared overlap integrals $ |{\mathcal{M} }_{if}|^2$ and $ |{\mathcal{E} }_{if}|^2$, respectively. The overlap integrals $ {\mathcal{M} }_{if}$ and $ {\mathcal{E} }_{if}$ can be expanded in $E_\gamma r$,  and generate   magnetic and electric multipole moments,
 \begin{eqnarray}
 {\mathcal{M} }_{if}&=&\int_0^\infty dr r^2 R_{nl}(r)R_{n^\prime l^\prime}^{\prime}(r)j_0\left(\frac{E_\gamma r}{2}\right)\nonumber\\
 &=&\delta_{nn^{\prime}} +\cdots \, , \nonumber \\
  {\mathcal{E} }_{if}&=&\frac{3}{E_\gamma}\int_0^\infty dr r^2 R_{nl}(r)R_{n^\prime l^\prime}^{\prime}(r)\left[\frac{E_\gamma r}{2}j_0\left(\frac{E_\gamma r}{2}\right)-j_1\left(\frac{E_\gamma r}{2}\right)\right]\nonumber\\
  &=&\int_0^\infty dr r^3 R_{nl}(r)R_{n^\prime l^\prime}^{\prime}(r)+\cdots \, ,
 \end{eqnarray}
 where the $R_{nl}$ are  radial wave-functions,  the $j_n$ are spherical Bessel functions, the $E_\gamma$ is the energy of the photons. Compared to the $M1$ transitions,  the decay widths of the $E1$ transitions maybe less sensitive to the radial wave-functions, therefore maybe  less sensitive to the LCWF.
 In the  phenomenological CLFQM, we usually  take  the three-dimensional harmonic-oscillator wave-functions in momentum space to approximate the radial wave-functions \cite{Cheng2003,Wang1002,WF-1}, jus like in the nonrelativistic quark models.

In this article, we intend to study the radiative transitions (the $E1$ transitions) among the vector and scalar heavy quarkonium states using the CLFQM as the first step, because  they can be taken as the sub-processes of the doubly radiative decays, and the calculations are relatively  simple.  Experimentally, the  widths of the radiative decays $\psi^{\prime} \to \chi_{c0}\,\gamma$, $\chi_{c0} \to J/\psi\,\gamma$, $\Upsilon^{\prime} \to \chi_{b0}\,\gamma$, $\Upsilon^{\prime\prime} \to \chi_{b0}\,\gamma$, $\Upsilon^{\prime\prime} \to \chi_{b0}^{\prime}\,\gamma$  have been precisely measured \cite{PDG}.
A large amount of bottomonium states  will be produced at the  Large Hadron Collider, and the radiative transitions will be studied experimentally.
We can study the radiative transitions by measuring the energy spectrum of the photons or reconstructing the final quarkonium states, although the soft
photons are difficult to identify.  The four-photon decay cascades   $\Upsilon(3 S)\to \gamma \chi_{bj}(2 P)\to \gamma\gamma\Upsilon(1 D)\to \gamma\gamma\gamma \chi_{bj}(1 P)\to \gamma\gamma\gamma\gamma\Upsilon(1 S)\to\gamma\gamma\gamma\gamma \ell^+\ell^-$ ($j=0,1,2$) have  been observed by the CLEO
collaboration \cite{CLEO-4}, where the softest photons have the energy less than $90\,\rm{MeV}$.

The article is arranged as follows:  we calculate the radiative transitions among the vector and scalar heavy quarkonium states
with the CLFQM in Sect.2;
in Sect.3, we present the numerical results and discussions; and Sect.4 is reserved for our
conclusions.

\section{Radiative decays with covariant light-front quark model}

The radiative transitions among the heavy quarkonium states can be
described by the following electromagnetic lagrangian $\mathcal{L}$,
\begin{eqnarray}
\mathcal{L}&=&-e e_b \bar{b}\gamma_\mu b A^\mu-e e_c \bar{c}\gamma_\mu c A^\mu \, ,
\end{eqnarray}
where the $A_\mu$ is the electromagnetic field,  the $e$ is the electromagnetic coupling constant, $e_b=-\frac{1}{3}$, and $e_c=\frac{2}{3}$. The transition amplitudes $\xi^\mu {\cal T}_{\mu}^{V\to S\gamma}$ and $\xi^\mu {\cal A}_{\mu}^{S\to V\gamma}$ can be decomposed as
\begin{eqnarray}
 {\cal T}_{\mu}^{V\to S\gamma}&=&  f_{VS\gamma}\left( \epsilon_\mu q \cdot P^{\prime}-P^{\prime}_\mu \epsilon \cdot q \right) \, , \nonumber \\
   {\cal A}_{\mu}^{S\to V\gamma}&=& g_{SV\gamma}\left( \epsilon^*_\mu q \cdot P-P_\mu \epsilon^* \cdot q \right) \, ,
\end{eqnarray}
according to  Lorentz covariance,
where  the $V$ and $S$ denote the vector and scalar mesons respectively,  the $P$, $P^{\prime}$, $q$ are the momenta of the initial mesons, final mesons and photons respectively, the $\xi_\mu$ and $\epsilon_\mu$ are  the polarization  vectors of the photons and vector mesons respectively,  the $f_{VS\gamma}$ and $g_{SV\gamma}$ are the electromagnetic form-factors (or the coupling constants) at $q^2=0$.  The amplitudes ${\cal T}_{\mu}^{V\to S\gamma} $ and ${\cal A}_{\mu}^{S\to V\gamma} $ satisfy conservation of electromagnetic currents. We can replace the photon's polarization  vector $\xi_\mu$  with its momentum $q_\mu$  in the transition amplitudes $\xi^\mu {\cal T}_{\mu}^{V\to S\gamma}$ and $\xi^\mu {\cal A}_{\mu}^{S\to V\gamma}$ to obtain $q^\mu {\cal T}_{\mu}^{V\to S\gamma} =q^\mu {\cal A}_{\mu}^{S\to V\gamma} =0$.

\begin{figure}
 \centering
 \includegraphics[totalheight=4cm,width=14cm]{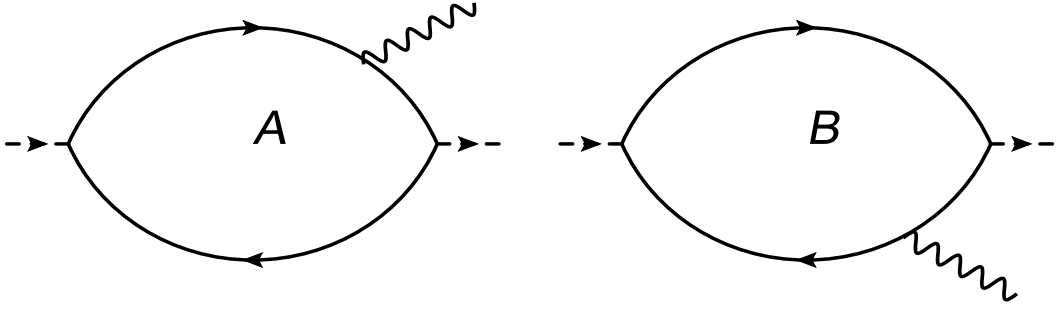}
    \caption{The Feynman diagrams contribute to the form-factors.  The photon is emitted from the quark (antiquark) line in the diagram $A$ ($B$). }
\end{figure}

From the lagrangian $\mathcal{L}$,  we can draw the Feynman diagrams (see Fig.1) and write down  the transition amplitudes,
\begin{eqnarray}
 {\cal T}_{\mu}^{V\to S\gamma} &=& -iee_QN_c \int\frac{d^4p_1}{(2\pi)^4}\left[\frac{H_V H_S}{N_1N_2N_1'}s_{\mu\alpha}^A+\frac{H_V H_S}{N_1N_2N_2'}s_{\mu\alpha}^B\right]\epsilon^{\alpha}_{V} \, ,  \\
 {\cal A}_{\mu}^{S\to V\gamma}&=& -iee_QN_c \int\frac{d^4p_1}{(2\pi)^4}\left[\frac{H_S H_V}{N_1N_2N_1'}\widetilde{s}_{\mu\alpha}^A+\frac{H_S H_V}{N_1N_2N_2'}\widetilde{s}_{\mu\alpha}^B\right]\epsilon^{*\alpha}_{V} \, ,
\end{eqnarray}
where $N_i= p_i^{2}-m_Q^2$, $N_i'= p_i'^{2}-m_Q^2$, and
\begin{eqnarray}
 s_{\mu\alpha}^A&=&{\rm Tr}\left\{(p\!\!\!\slash'_1+m_Q)\gamma_\mu(p\!\!\!\slash_1+m_Q)\left[\gamma_\alpha-\frac{(p_1-p_2)_\alpha}{W_V}\right](-p\!\!\!\slash_2+m_Q) \right\} \, , \nonumber \\
 s_{\mu\alpha}^B&=&{\rm Tr}\left\{(p\!\!\!\slash_1+m_Q)\left[\gamma_\alpha-\frac{(p_1-p_2)_\alpha}{W_V}\right](-p\!\!\!\slash_2+m_Q) \gamma_\mu(-p\!\!\!\slash'_2+m_Q) \right\} \, , \\
 \widetilde{s}_{\mu\alpha}^A&=&{\rm Tr}\left\{(p\!\!\!\slash'_1+m_Q)\gamma_\mu(p\!\!\!\slash_1+m_Q)(-p\!\!\!\slash_2+m_Q)\left[\gamma_\alpha-\frac{(p^{\prime}_1-p_2)_\alpha}{W_V}\right] \right\} \, , \nonumber \\
 \widetilde{s}_{\mu\alpha}^B&=&{\rm Tr}\left\{(p\!\!\!\slash_1+m_Q)(-p\!\!\!\slash_2+m_Q) \gamma_\mu(-p\!\!\!\slash'_2+m_Q) \left[\gamma_\alpha-\frac{(p_1-p^{\prime}_2)_\alpha}{W_V}\right]\right\} \, ,
\end{eqnarray}
 the $H_V$ and $H_S$  are the LFWF of the  vector and  scalar mesons respectively, the $A$ and $B$ denote the diagrams in which the photon emitted from the quark and antiquark lines, respectively (see Fig.1), and the $W_V$ is a parameter. In writing the transition amplitudes ${\cal T}_{\mu}^{V\to S\gamma}$ and ${\cal A}_{\mu}^{S\to V\gamma}$, we have used the following definitions of
 the quark-meson-antiquark vertexes $i\Gamma_M$ \cite{Cheng2003},
\begin{eqnarray}
i\Gamma_M&=&-iH_S  \,\,\,\,\,\,\,\,\,\,\,\,\,\,\,\,\,\,\,\,\,\,\,\,\,\,\,\,\,\,\,\,\,\,\,\,\,\,\,\,\,\,\,\,\,\,\,\,\,\rm{for\,\, scalar\,\, meson} \, ,\nonumber\\
         &=&iH_V\left[\gamma_\mu-\frac{(p_1-p_2)_\mu}{W_V} \right]\,\,\,\rm{for\,\, vector\,\, meson} \, ,
\end{eqnarray}
where the $p_1$ and $p_2$ are momenta of the quark and antiquark, respectively.

 We assume that the covariant LFWF $H_V$ and $H_S$  are analytic in the upper (or lower) $p_1^-$ complex plane, close  the integral contour in the upper (or lower) $p_1^-$ complex plane for the diagram $A$ (or $B$),  which corresponds to set the
antiquark (or quark) on the mass-shell. The one-shell restrictions  are implemented by the following replacements,
\begin{eqnarray}
p_2&\to&\hat{p}_2^2=m_Q^2\, , \, \, \,p_1\to\hat{p}_1=P-\hat{p}_2\, , \nonumber\\
N_1 &\to& \hat{N}_1=\hat{p}_1^2-m_Q^2\, , \, \, \,N^{\prime}_1 \to \hat{N}^{\prime}_1=\hat{p}_1^{\prime2}-m_Q^2\, , \nonumber\\
H_V &\to& \hat{H}_V=h_V(x_2,p_\perp)\, , \, \, \,H_S \to \hat{H}_S=h_S(x_2,p_\perp)\, , \, \, \,W_V \to \hat{W}_V=w_V\, , \nonumber\\
\int\frac{d^4p_1}{N_1N_2N_1'}&\to&-i\pi\int \frac{dx_2d^2p_{\perp}}{x_2\hat{N}_1\hat{N}^{\prime}_1}\, ,
\end{eqnarray}
for the Feynman diagram $A$ and
\begin{eqnarray}
p_1&\to&\hat{p}_1^2=m_Q^2\, , \, \, \, p_2\to\hat{p}_2=P-\hat{p}_1\, , \nonumber\\
N_2 &\to& \hat{N}_2=\hat{p}_2^2-m_Q^2\, , \, \, \,
N^{\prime}_2 \to \hat{N}^{\prime}_2=\hat{p}_2^{\prime2}-m_Q^2  \, , \nonumber\\
H_V &\to& \hat{H}_V=h_V(x_2,p_\perp)\, , \, \, \,H_S \to \hat{H}_S=h_S(x_2,p_\perp)\, , \, \, \,W_V \to \hat{W}_V=w_V\, , \nonumber\\
\int\frac{d^4p_1}{N_1N_2N_2^{\prime}}&\to&-i\pi\int \frac{dx_2d^2p_{\perp}}{x_1\hat{N}_2\hat{N}^{\prime}_2}\, ,
\end{eqnarray}
for the Feynman diagram $B$, where we have used the light-front decomposition of the momenta $P=(P^+,P^-,0_\perp)$, $P^{\pm}=P^0 \pm P^3$, $p^+_{1,2}=x_{1,2}P^+$, $x_1+x_2=1$, $p_{1,2\perp}=\pm p_\perp$, $q=(0,q^-,q_\perp)=(0,q^-,0_\perp)$. Finally, we obtain the form-factors or coupling constants   at zero momentum transition\footnote{For technical details, one can consult Refs.\cite{Jaus1999,Jaus2003,Cheng2003}.},
\begin{eqnarray}
f_{VS\gamma}&=&\frac{e_Q e N_c}{8\pi^3}\int dx_2 d^2p_\perp \frac{h_V(x_2,p_\perp) h_S(x_2,p_{\perp})}{x_1^2 x_2(M_i^2-M_0^2)(M_f^2-M_{0}^2) } \nonumber\\
&&\left\{m_Q(M_i^2-M_f^2)-2m_Qx_1(M_i^2-M_0^2)-\left[4m_Q+\frac{x_1(M_i^2+M_f^2)+2p_\perp^2-8x_1m_Q^2}{x_1w_V} \right]p_\perp^2\right\}  \nonumber\\
&&+\frac{e_Q e N_c}{8\pi^3}\int dx_2 d^2p_\perp \frac{h_V(x_2,p_\perp) h_S(x_2,p_{\perp})}{x_1  x_2^2(M_i^2-M_0^2)(M_f^2-M_{0}^2) } \nonumber\\
&&\left\{m_Q(M_i^2-M_f^2)-2m_Qx_2(M_i^2-M_0^2)-\left[4m_Q+\frac{x_2(M_i^2+M_f^2)+2p_\perp^2-8x_2m_Q^2}{x_2w_V} \right]p_\perp^2\right\} \, , \nonumber\\
\end{eqnarray}
\begin{eqnarray}
g_{SV\gamma}&=&\frac{e_Q e N_c}{8\pi^3}\int dx_2 d^2p_\perp \frac{h_V(x_2,p_\perp) h_S(x_2,p_{\perp})}{x_1^2 x_2(M_i^2-M_0^2)(M_f^2-M_{0}^2) } \nonumber\\
&&\left\{m_Q(M_f^2-M_i^2)-2m_Qx_1(M_f^2-M_0^2)-\left[4m_Q+\frac{x_1(M_i^2+M_f^2)+2p_\perp^2-8x_1m_Q^2}{x_1w_V} \right]p_\perp^2\right\}  \nonumber\\
&&+\frac{e_Q e N_c}{8\pi^3}\int dx_2 d^2p_\perp \frac{h_V(x_2,p_\perp) h_S(x_2,p_{\perp})}{x_1  x_2^2(M_i^2-M_0^2)(M_f^2-M_{0}^2) } \nonumber\\
&&\left\{m_Q(M_f^2-M_i^2)-2m_Qx_2(M_f^2-M_0^2)-\left[4m_Q+\frac{x_2(M_i^2+M_f^2)+2p_\perp^2-8x_2m_Q^2}{x_2w_V} \right]p_\perp^2\right\} \, , \nonumber\\
\end{eqnarray}
where the LFWF are defined as \cite{Cheng2003},
\begin{eqnarray}
 h_V(x_2,p_\perp)&=& (M_{i/f}^2-M_0^2)\sqrt{\frac{x_1x_2}{N_c}}\frac{1}{{\sqrt 2} M_0}\phi_S(x_2,p_\perp)\, ,\nonumber\\
 h_S(x_2,p_\perp)&=& (M_{i/f}^2-M_0^2)\sqrt{\frac{x_1x_2}{N_c}}\frac{1}{2{\sqrt 6}  }\phi_P(x_2,p_\perp)\, ,\nonumber\\
 \phi_P(x_2,p_\perp)&=&\sqrt{\frac{2}{\beta^2}}\phi_S(x_2,p_\perp) \, , \nonumber\\
   \phi_{1S}(x_2,p_\perp)&=& 4\left(\frac{\pi}{\beta^2}\right)^{3/4}
 \sqrt{\frac{dp_z}{dx_2}}{\exp}\left(-\frac{p_{\bot}^2+p_z^2}{2\beta^2}\right) \, ,\nonumber\\
 \phi_{2S}(x_2,p_\perp)&=& 4\sqrt
 {\frac{2}{3}}\left(\frac{\pi}{\beta^2}\right)^{3/4}
 \sqrt{\frac{dp_z}{dx_2}}{\exp}\left(-\frac{p_{\bot}^2+p_z^2}{2\beta^2}\right)  \left[\frac{p_{\bot}^2+p_z^2}{\beta^2}-\frac{3}{2}\right]\, ,\nonumber\\
 \phi_{3S}(x_2,p_\perp)&=&  4\sqrt
 {\frac{2}{15}}\left(\frac{\pi}{\beta^2}\right)^{3/4}  \sqrt{\frac{dp_z}{dx_2}}{\exp}\left(-\frac{p_{\bot}^2+p_z^2}{2\beta^2}\right) \left[\frac{(p_{\bot}^2+p_z^2)^2}{\beta^4}-\frac{5(p_{\bot}^2+p_z^2)}{\beta^2}+\frac{15}{4}\right] \, , \nonumber\\
  \frac{dp_z}{dx_2}&=&\frac{e_1 e_2}{x_1x_2M_0}\,  ,  \,  w_V=M_0+2m_Q \,,\nonumber\\
 M^{2}_0 &=&(e_1+e_2)^2=\frac{p^{2}_\perp+m_Q^{2}}{x_1}+\frac{p^{2}_{\perp}+m_Q^2}{x_2}\, , \nonumber\\
 p_z&=&\frac{x_2 M_0}{2}-\frac{m_Q^2+p^{2}_\bot}{2 x_2 M_0}\,  , \, e_i =\sqrt{m^{2}_Q+p^{2}_\bot+p^{2}_z} \, ,
 \end{eqnarray}
the $M_i$ and $M_f$ are the masses of the  initial and final heavy quarkonium states, the $e_i$  can be viewed as the energy of the quark (antiquark),
and  the  $M_0$ can be viewed as the kinetic invariant mass of the quark-antiquark system.

In calculations, we have used  the following rules \cite{Jaus1999,Jaus2003,Cheng2003},
 \begin{eqnarray}
\hat{p}^\mu_1 &\doteq& P^\mu A_1^{(1)}+q^\mu A_2^{(1)} \, , \nonumber \\
\hat{p}^\mu_1 \hat{p}^\nu_1 &\doteq& g^{\mu\nu} A_1^{(2 )}+P^\mu P^\nu A_2^{(2)}+(P^\mu q^\nu+q^\mu P^\nu) A_3^{(2)}+q^\mu q^\nu A_4^{(2)} \, , \nonumber \\
\hat{N}_2 &\to & Z_2 \, ,\nonumber \\
\int d^2 p_\perp \,\frac{(p_\perp \cdot q_\perp)^2}{q^2}&=&-\frac{1}{2}\int d^2 p_\perp \, p_\perp^2 \, , \,
\int dx_2 d^2p_\perp \frac{h_V h_S}{ x_2 \hat{N}_1 \hat{N}_1^{\prime} }\left(x_1 Z_2-2A_1^{(2)}\right) =0 \, , \nonumber \\
A_1^{(1)}&=&\frac{x_1}{2}\,  ,  \, A_2^{(1)}=\frac{x_1}{2}-\frac{p_\perp \cdot q_\perp}{q^2} \, , \, A_1^{(2)}=-p_\perp^2-\frac{(p_\perp \cdot q_\perp)^2}{q^2} \, ,\nonumber \\
A_{2}^{(2)}&=& A_{1}^{(1)}A_{1}^{(1)} \, , \, A_{3}^{(2)}= A_{1}^{(1)}A_{2}^{(1)} \, , \, A_{4}^{(2)}= A_{2}^{(1)}A_{2}^{(1)}-\frac{A_1^{(2)}}{q^2} \, .
 \end{eqnarray}
In going from the manifestly covariant Feynman integral to the
light-front one, there appear additional spurious contributions
proportional to the lightlike vector $\omega^\mu=(1,0,0,-1)$. The additional residual
contributions are expressed in terms of the
$B^{(m)}_{n}$ and $C^{(m)}_{n}$ functions \cite{Jaus1999,Jaus2003}. The $C^{(m)}_{n}$ functions are  canceled if the zero mode contributions are correctly taken
 into account, while the
 $B^{(m)}_{n}$ functions under integration vanish or are numerically tiny \cite{Jaus1999,Jaus2003,Cheng2003}. The  rules in Eq.(14) have accounted for the zero mode contributions.  The results are covariant and free of
 spurious contributions.

\section{Numerical results and discussions}
 We take the masses of the heavy quarkonium states from the Particle Data Group, $M_{\chi_{c0}}=3414.75\,\rm{MeV}$,  $M_{J/\psi}=3096.916\,\rm{MeV}$, $M_{\psi^{\prime}}=3686.09\,\rm{MeV}$, $M_{\Upsilon}=9.46030\,\rm{GeV}$, $M_{\Upsilon^{\prime}}=10.02326\,\rm{GeV}$, $M_{\Upsilon^{\prime\prime}}=10.3552\,\rm{GeV}$,
 $M_{\chi_{b0}}=9.85944\,\rm{GeV}$, $M_{\chi^{\prime}_{b0}}=10.2325\,\rm{GeV}$  \cite{PDG}, and
 obtain the radiative decay widths,
  \begin{eqnarray}
  \Gamma_{\chi_{c0}\to J/\psi \gamma}&=&2214.63\,e_c^2e^2\,g_{\chi_{c0} J/\psi \gamma}^2 \,\rm{KeV} \, , \nonumber \\
   \Gamma_{\psi^{\prime}\to \chi_{c0} \gamma}&=&473.535\,e_c^2e^2\,f_{\psi^{\prime} \chi_{c0} \gamma}^2 \,\rm{KeV} \, , \nonumber \\
   \Gamma_{\Upsilon^{\prime}\to \chi_{b0} \gamma}&=&113.783\,e_b^2e^2\,f_{\Upsilon^{\prime} \chi_{b0} \gamma}^2 \,\rm{KeV} \, , \nonumber \\
   \Gamma_{\Upsilon^{\prime\prime}\to \chi_{b0} \gamma}&=&3005.49\,e_b^2e^2\,f_{\Upsilon^{\prime\prime} \chi_{b0} \gamma}^2 \,\rm{KeV} \, , \nonumber \\
   \Gamma_{\Upsilon^{\prime\prime}\to \chi^{\prime}_{b0} \gamma}&=&48.135\,e_b^2e^2 \,f_{\Upsilon^{\prime\prime} \chi^{\prime}_{b0} \gamma}^2 \,\rm{KeV} \, .
   \end{eqnarray}
   The numerical factors in front of the coupling constants have hierarchies as the decay widths proportional to $E^3_\gamma$,
   the energy of the photons $E_\gamma=\frac{M_i^2-M_f^2}{2M_i}$.

  Compared to the experimental data from the Particle Data Group \cite{PDG},
  \begin{eqnarray}
 \Gamma_{\chi_{c0}\to J/\psi \gamma}&=&122.85\pm 9.36 \pm 8.4 \,\, (\rm or\,\, 121.68\pm7.02\pm 8.32)\,\rm{KeV} \, , \nonumber \\
\Gamma_{\psi^{\prime}\to \chi_{c0} \gamma}&=&27.6848\pm 1.5488 \pm 0.8866 \, \, (\rm or \,\, 29.4272\pm 0.8712 \pm 0.9424)\,\rm{KeV} \, ,  \\
\Gamma_{\Upsilon^{\prime}\to \chi_{b0} \gamma}&=&1.21524\pm0.09994\pm0.12792\,\rm{KeV} \, ,\nonumber \\
\Gamma_{\Upsilon^{\prime\prime}\to \chi_{b0} \gamma}&=&0.06096\pm0.004995 \pm 0.008128\,\rm{KeV} \, , \nonumber \\
\Gamma_{\Upsilon^{\prime\prime}\to \chi_{b0}^{\prime} \gamma}&=&1.19888\pm0.10915 \pm 0.12192\,\rm{KeV}  \, ,
\end{eqnarray}
where the  first uncertainties come from the total widths and the second uncertainties come from the branching ratios.  The radiative widths $\Gamma_{\chi_{c0}\to J/\psi \gamma}$ and $\Gamma_{\psi^{\prime}\to \chi_{c0} \gamma}$ in Eq.(16) come from the  Particle Data Group's  average (or fitted) values of the total widths $\Gamma_{\chi_{c0}}$ and $\Gamma_{\psi^{\prime}}$. From Eqs.(15-17), we can draw the conclusion tentatively that the coupling constants $f^2_{VS\gamma}$  differ greatly from each other for the bottomonium states, while $g_{\chi_{c0} J/\psi \gamma}^2\approx f_{\psi^{\prime} \chi_{c0} \gamma}^2$.

 From Eqs.(11-13), we can see that the form-factors or coupling constants depend on
two kinds of inputs parameters,   the constituent quark masses $m_Q$ and the shape parameters  $\beta$ of the
LFWF.   The spin averaged ground state masses are $\frac{M_{\eta_c}+3M_{J/\psi}}{4}=3.068\,\rm{GeV}$,
$\frac{M_{\eta_b}+3M_{\Upsilon}}{4}=9.443\,\rm{GeV}$ from the Particle Data Group \cite{PDG}, we  estimate the constituent quark masses $m_c\approx1.5\,\rm{GeV}$ and $m_b\approx4.7\,\rm{GeV}$. In this article, we take the constituent quark masses $m_Q$ and shape parameters $\beta$ of the LFWF as free parameters and search for the optimal values at the ranges $m_c=(1.3-1.7)\,\rm{GeV}$ and $m_b=(4.5-5.0)\,\rm{GeV}$.

In numerical calculations, we observe that the form-factors (or coupling constants) are sensitive to the constituent quark masses $m_Q$ and the shape parameters  $\beta$, small  variations of those input parameters can lead to large changes of the form-factors (thereafter the decay widths). The optimal values are $m_c=1.5\,\rm{GeV}$ and $m_b=4.8\,\rm{GeV}$, $\beta_{\chi_{c0}}=1.30\,\rm{GeV}$, $\beta_{J/\psi}=0.76\,\rm{GeV}$,
$\beta_{\psi^{\prime}}=0.83\,\rm{GeV}$, $\beta_{\Upsilon^{\prime}}=1.30\,\rm{GeV}$, $\beta_{\Upsilon^{\prime\prime}}=0.976\,\rm{GeV}$,
$\beta_{\chi_{b0}}=0.94\,\rm{GeV}$, $\beta_{\chi_{b0}^{\prime}}=0.945\,\rm{GeV}$, which can reproduce the decay widths of the five observed processes  $\chi_{c0}\to J/\psi \gamma$, $\psi^{\prime}\to \chi_{c0} \gamma$, $\Upsilon^{\prime}\to \chi_{b0} \gamma$, $\Upsilon^{\prime\prime}\to \chi_{b0} \gamma$ and $\Upsilon^{\prime\prime}\to \chi_{b0}^{\prime} \gamma$ \cite{PDG}.
If the form-factors (or coupling constants) are not sensitive to the shape parameters $\beta$ of the LFWF, we can introduce two parameters to characterize the charmonium  and  bottomonium states respectively, and search for the optimal values to reproduce the decay widths of the five observed radiative transitions. However, two parameters cannot lead to satisfactory results.

The optimal values  $m_c=1.5\,\rm{GeV}$ and $m_b=4.8\,\rm{GeV}$ approximate or equal to the
estimated values  $m_c\approx1.5\,\rm{GeV}$ and $m_b\approx4.7\,\rm{GeV}$.  In the constituent quark model, the usually used constituent quark masses are $m_c=(1.4-1.6)\,\rm{GeV}$ and $m_b=(4.7-4.9)\,\rm{GeV}$, it is natural to take the values
 $m_c=1.5\pm0.1\,\rm{GeV}$ and $m_b=4.8\pm0.1\,\rm{GeV}$ in this article.

 We choose (or suppose) the shape parameters have the same uncertainties as the constituent quark masses tentatively, i.e. $\beta_{\chi_{c0}}=1.30\pm0.1\,\rm{GeV}$, $\beta_{J/\psi}=0.76\pm0.1\,\rm{GeV}$,
$\beta_{\psi^{\prime}}=0.83\pm0.1\,\rm{GeV}$, $\beta_{\Upsilon^{\prime}}=1.30\pm0.1\,\rm{GeV}$, $\beta_{\Upsilon^{\prime\prime}}=0.976\pm0.1\,\rm{GeV}$,
$\beta_{\chi_{b0}}=0.94\pm0.1\,\rm{GeV}$, $\beta_{\chi_{b0}^{\prime}}=0.945\pm0.1\,\rm{GeV}$. The resulting radiative decay widths are
\begin{eqnarray}
 \Gamma_{\chi_{c0}\to J/\psi \gamma}&=&121.54_{-97.18}^{+160.22}\,{}_{-26.26}^{+34.32}\,{}_{-32.67}^{+39.89}\,\rm{KeV} \, , \nonumber \\
\Gamma_{\psi^{\prime}\to \chi_{c0} \gamma}&=&27.90_{-17.97}^{+26.67}\,{}^{+9.83}_{-9.10} \,{}^{+1.56}_{-3.55}\,\rm{KeV} \, , \nonumber \\
\Gamma_{\Upsilon^{\prime}\to \chi_{b0} \gamma}&=&1.23^{+2.31}_{-1.13}\,{}^{+0.07}_{-0.13} \,{}^{+0.52}_{-0.34}\,\rm{KeV} \, ,\nonumber \\
\Gamma_{\Upsilon^{\prime\prime}\to \chi_{b0} \gamma}&=&0.0606^{+0.0078}_{-0.0072}\,{}^{+0.2160}_{-0.0605} \,{}^{+0.2343}_{-0.0589}\,\rm{KeV} \, , \nonumber \\
\Gamma_{\Upsilon^{\prime\prime}\to \chi_{b0}^{\prime} \gamma}&=&1.19^{+0.31}_{-0.27}\,{}^{+2.30}_{-1.18} \,{}^{+3.75}_{-1.14}\,\rm{KeV}  \, ,
\end{eqnarray}
where the uncertainties come from the heavy quark masses and shape parameters of  the initial and final quarkonium states, sequentially.

From Eqs.(16-18), we can see that  the central values of the present predictions are consistent with the experimental data from the Particle Data Group \cite{PDG}. However,  the uncertainties come from the heavy quark masses exceed $100\%$ for the transitions
$\chi_{c0}\to J/\psi \gamma$, $\psi^{\prime}\to \chi_{c0} \gamma$, $\Upsilon^{\prime}\to \chi_{b0} \gamma$, and the   uncertainties come from the shape parameters
also exceed $100\%$ for the transitions $\Upsilon^{\prime\prime}\to \chi_{b0} \gamma$, $\Upsilon^{\prime\prime}\to \chi_{b0}^{\prime} \gamma$. We have little room   for varying those parameters to reproduce the experimental data, which weakens the predictive ability of the CLFQM remarkably. On the other hand, if we take  the uncertainties of the experimental data on the radiative transitions (see Eqs.(16-17)) to constrain  the uncertainties of the constituent quark masses and shape parameters, the allowed uncertainties are $|\delta m_c|< 10\,\rm{MeV}$, $|\delta m_b|< 10\,\rm{MeV}$, $|\delta\beta_{\chi_{c0}}|<35\,\rm{MeV}$,  $|\delta\beta_{J/\psi}|<30\,\rm{MeV}$,
$|\delta\beta_{\psi^{\prime}}|<17\,\rm{MeV}$, $|\delta\beta_{\Upsilon^{\prime}}|<100\,\rm{MeV}$, $|\delta\beta_{\Upsilon^{\prime\prime}}|<8\,\rm{MeV}$, $|\delta\beta_{\chi_{b0}}|<10\,\rm{MeV}$, $|\delta \beta_{\chi_{b0}^{\prime}}|<6\,\rm{MeV}$.

The heavy quarkonium states have equal constituent quark masses, $m_{Q}=m_{\overline{Q}}$, the heavy quark masses  are taken as one parameter $m_Q$ rather than two parameters $m_{Q}$ and $m_{\overline{Q}}$, see Eqs.(11-13),  the total uncertainties come from the heavy quark masses are larger than that come from the heavy quark masses $m_Q$ and $m_{\overline{Q}}$, respectively.

In Figs.2-3, we plot the radiative decay widths $\Gamma_{\chi_{c0}\to J/\psi \gamma}$, $\Gamma_{\psi^{\prime}\to \chi_{c0} \gamma}$, $\Gamma_{\Upsilon^{\prime}\to \chi_{b0} \gamma}$, $\Gamma_{\Upsilon^{\prime\prime}\to \chi_{b0} \gamma}$ and
 $\Gamma_{\Upsilon^{\prime\prime}\to \chi_{b0}^{\prime} \gamma}$ with variations of the shape parameters $\beta$.
From the figures, we can see that the widths are sensitive to the shape parameters $\beta$ indeed, small variations of the shape parameters $\beta$ can lead to rather large changes of the decay widths, especially when there are nodi in the radial wave-functions.  It is
very difficult (or impossible) to obtain two universal parameters $\beta_{\bar{c}c}$ and $\beta_{\bar{b}b}$ to characterize the charmonium  and  bottomonium states respectively.

We can take the  optimal values  $m_c=1.5\,\rm{GeV}$ and $m_b=4.8\,\rm{GeV}$, $\beta_{\chi_{c0}}=1.30\,\rm{GeV}$, $\beta_{J/\psi}=0.76\,\rm{GeV}$,
$\beta_{\psi^{\prime}}=0.83\,\rm{GeV}$, $\beta_{\Upsilon^{\prime}}=1.30\,\rm{GeV}$, $\beta_{\Upsilon^{\prime\prime}}=0.976\,\rm{GeV}$,
$\beta_{\chi_{b0}}=0.94\,\rm{GeV}$, $\beta_{\chi_{b0}^{\prime}}=0.945\,\rm{GeV}$ as the basic  input parameters and study other radiative transitions among the heavy quarkonium states. For example,
we can take the parameter $\beta_{\chi_{b0}}=0.94\,\rm{GeV}$ to calculate the decay width $\Gamma_{\chi_{b0}\to \Upsilon \gamma}$. From Fig.3, we can see that the prediction $\Gamma_{\chi_{b0}\to \Upsilon \gamma}=(18-34)\,\rm{KeV}$ for $\beta_{\Upsilon}=(0.4-1.4)\,\rm{GeV}$  is consistent with the value $\Gamma_{\chi_{b0}\to \Upsilon \gamma}=(24-26)\,\rm{KeV}$ from the potential quark models \cite{PQM-width}.
If we take the parameter $\beta_{\Upsilon}=1.16\,\rm{GeV}$, the predictions $\Gamma_{\chi_{b0}\to \Upsilon \gamma}=24.75\,\rm{KeV}$ and $\Gamma_{\chi_{b0}^{\prime}\to \Upsilon \gamma}=5.97\,\rm{KeV}$  are consistent with the values from the potential quark models $\Gamma_{\chi_{b0}\to \Upsilon \gamma}=(24-26)\,\rm{KeV}$ and $\Gamma_{\chi_{b0}^{\prime}\to \Upsilon \gamma}=(4.5-8.5)\,\rm{KeV}$ \cite{PQM-width}, while the prediction $\Gamma_{\chi_{b0}^{\prime}\to \Upsilon^{\prime} \gamma}=46.74\,\rm{KeV}$ is much larger than the value $\Gamma_{\chi_{b0}^{\prime}\to \Upsilon^{\prime} \gamma}=(11-12)\,\rm{KeV}$ from the  potential quark model \cite{PQM-width}.

 Without precise experimental data, we cannot determine
the value of the shape parameter $\beta_{\Upsilon}$.
On the other hand, there are controversies for the spectroscopy of the charmonium states, and lack experimental  data on other radiative transitions among the vector and scalar charmonium states. We prefer study those processes in the future.
In Ref.\cite{Wang1002}, the author calculates the $S$-wave to $S$-wave radiative  transitions $\Upsilon \to \eta_b \gamma$, $\Upsilon^{\prime} \to \eta_b \gamma$,
$\Upsilon^{\prime} \to \eta_b^{\prime} \gamma$,  $\Upsilon^{\prime\prime} \to \eta_b \gamma$,
$\Upsilon^{\prime\prime} \to \eta_b^{\prime} \gamma$, $J/\psi \to \eta_c\gamma$, $\psi^{\prime} \to \eta_c\gamma$, $\psi^{\prime} \to \eta_c^{\prime}\gamma$
with the CLFQM, and observes that those  $M1$ transitions are sensitive to the heavy quark masses and shape parameters of the LFWF, the existing experimental data cannot be reproduced consistently with suitable parameters $m_Q$ and $\beta$. In Ref.\cite{Ke-1006}, the authors  modify the LFWF by introducing several additional parameters besides the shape parameters $\beta$   and study the radiative decays $\Upsilon(nS)\to \eta_b+\gamma$, the prediction  ${\rm Br}\left(\Upsilon(3S)\to \eta_b+\gamma\right)=(1.87 \pm 0.71) \times 10^{-4}$ is much smaller than the experimental data $( 5.1 \pm 0.7 ) \times 10^{-4}$ \cite{PDG}.  We can draw the conclusion tentatively that we cannot get rid of  the sensitivity to the shape parameters $\beta$ without introducing several additional parameters. However, the experimental data are far from enough to fit the additional parameters.

\begin{figure}
 \centering
 \includegraphics[totalheight=5cm,width=6cm]{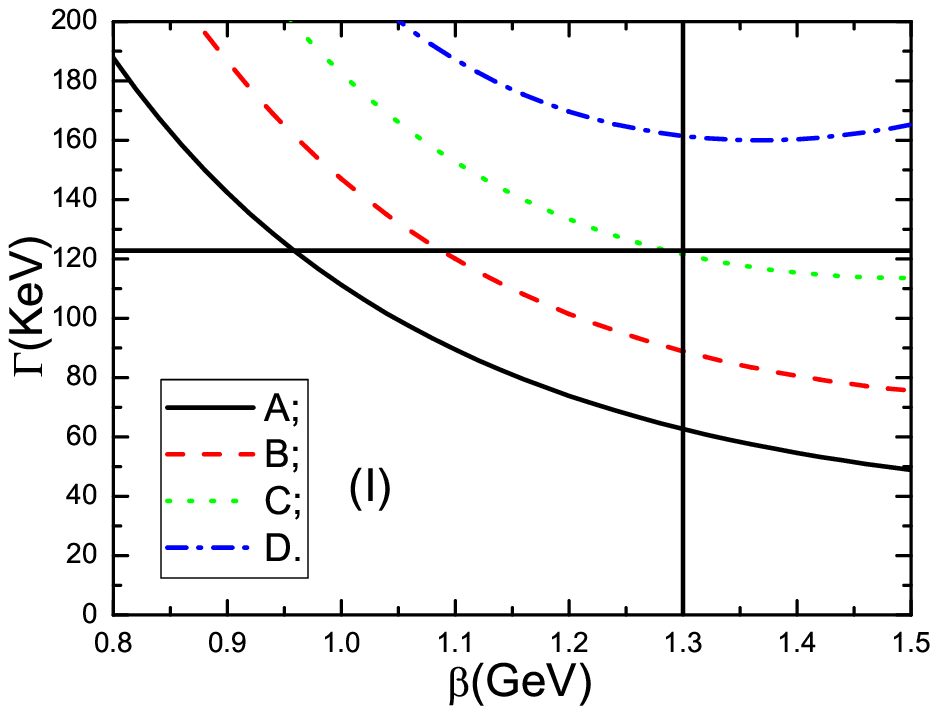}
 \includegraphics[totalheight=5cm,width=6cm]{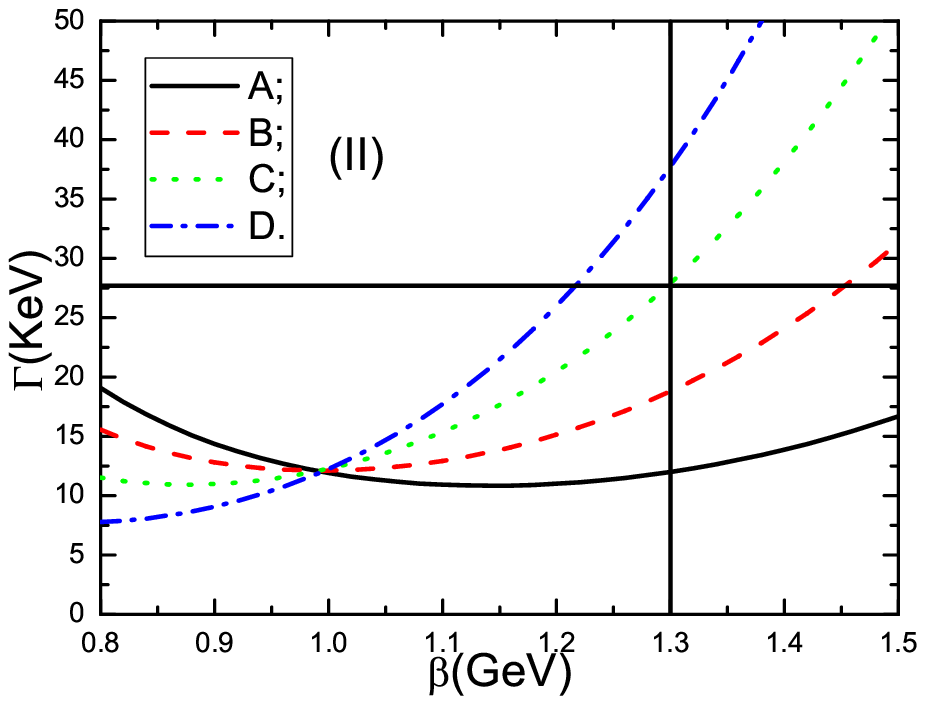}
        \caption{ The radiative decay widths with variations of the shape parameter $\beta_{\chi_{c0}}$. In (I) $\chi_{c0}\to J/\psi \gamma$, $A$, $B$, $C$, $D$ correspond to $\beta_{J/\psi}/\beta_{\chi_{c0}}=0.56/1.3$, $0.66/1.3$, $0.76/1.3$, $0.86/1.3$, respectively; in (II) $\psi^{\prime}\to \chi_{c0} \gamma$, $A$, $B$, $C$, $D$ correspond to $\beta_{\psi^{\prime}}/\beta_{\chi_{c0}}=0.63/1.3$, $0.73/1.3$, $0.83/1.3$, $0.93/1.3$, respectively.}
\end{figure}

\begin{figure}
 \centering
 \includegraphics[totalheight=5cm,width=6cm]{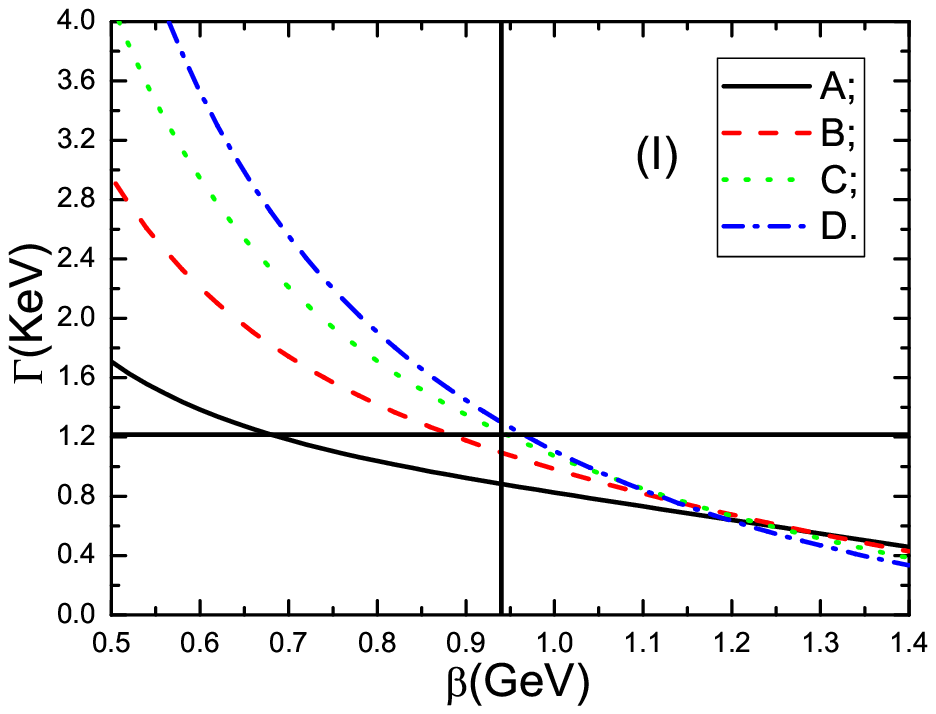}
 \includegraphics[totalheight=5cm,width=6cm]{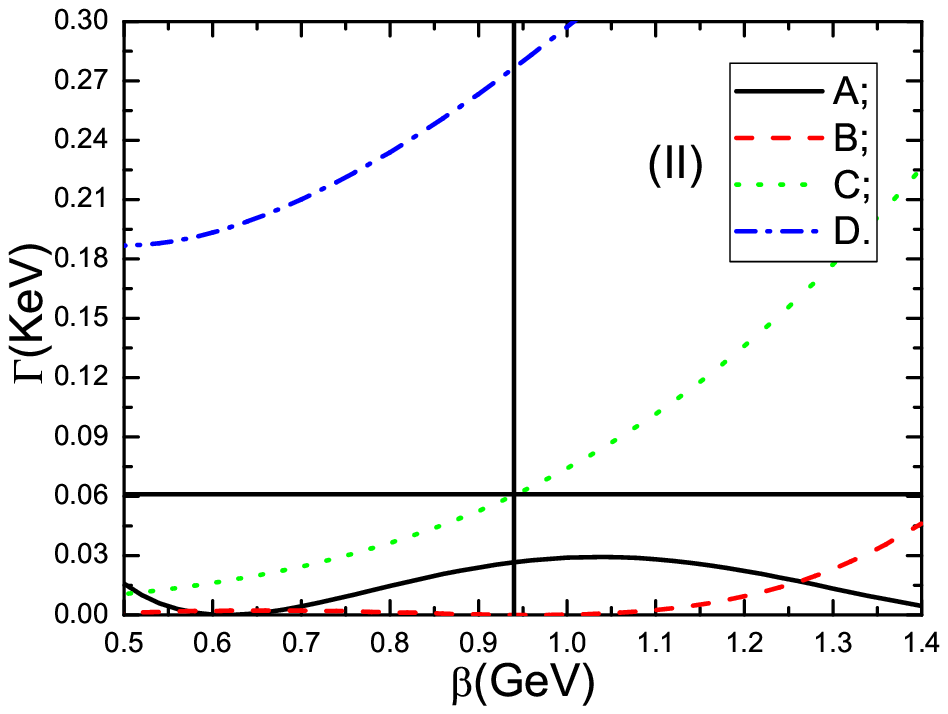}
 \includegraphics[totalheight=5cm,width=6cm]{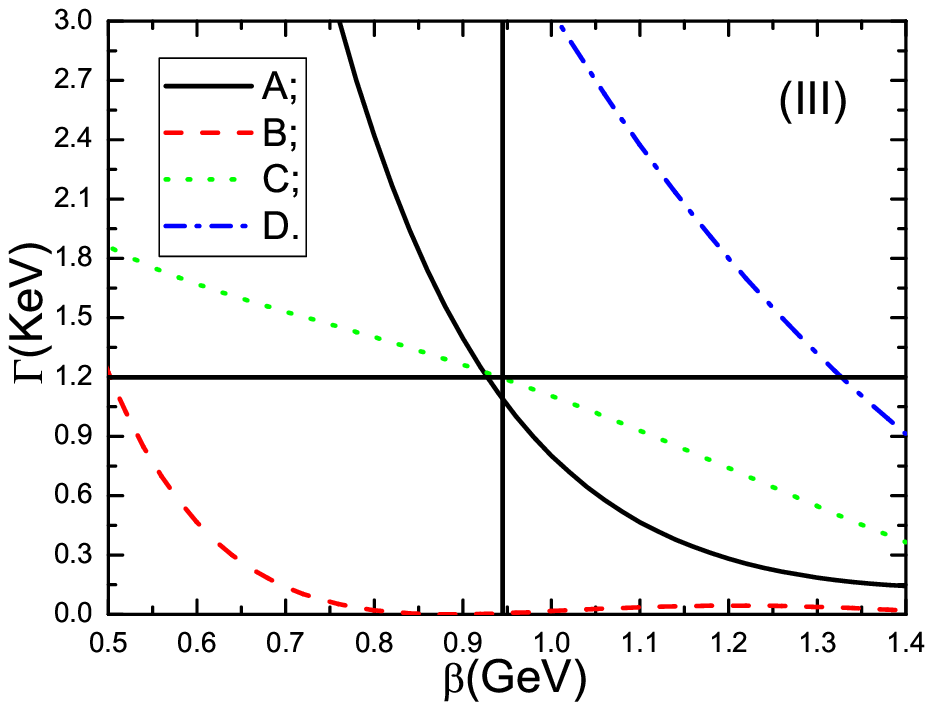}
 \includegraphics[totalheight=5cm,width=6cm]{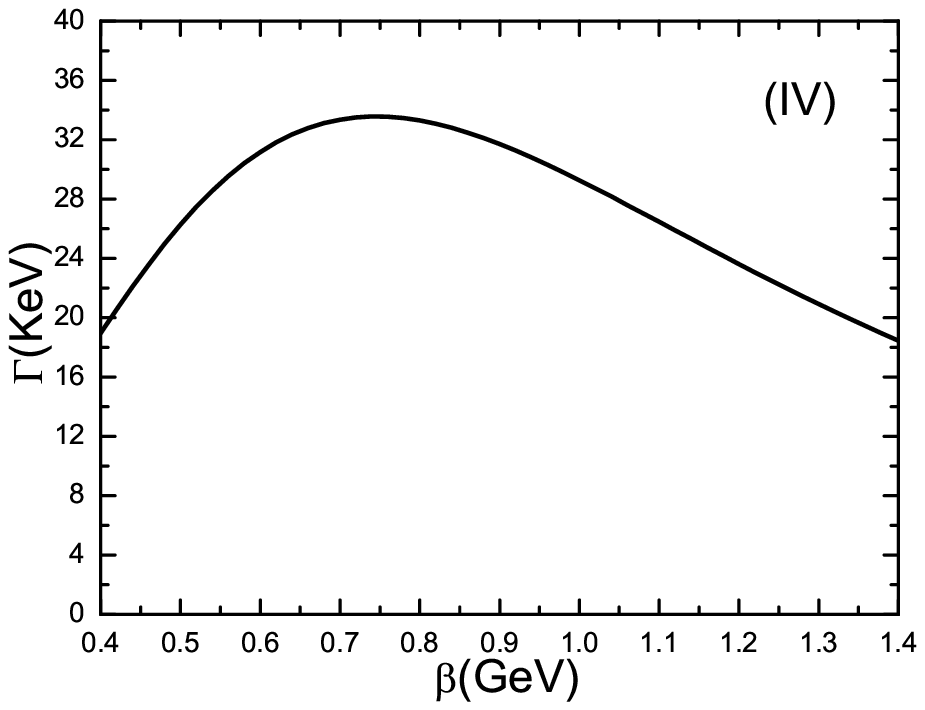}
        \caption{ The radiative decay widths with variations of the shape parameters $\beta_{\chi_{b0}}$ or $\beta_{\chi_{b0}^{\prime}}$. In (I) $\Upsilon^{\prime}\to \chi_{b0} \gamma$, $A$, $B$, $C$, $D$ correspond to $\beta_{\Upsilon^{\prime}}/\beta_{\chi_{b0}}=1.1/0.94$, $1.2/0.94$, $1.3/0.94$, $1.4/0.94$, respectively; in (II) $\Upsilon^{\prime\prime}\to \chi_{b0} \gamma$, $A$, $B$, $C$, $D$ correspond to $\beta_{\Upsilon^{\prime\prime}}/\beta_{\chi_{b0}}=0.776/0.94$, $0.876/0.94$, $0.976/0.94$, $1.076/0.94$, respectively; in (III) $\Upsilon^{\prime\prime}\to \chi_{b0}^{\prime} \gamma$, $A$, $B$, $C$, $D$ correspond to $\beta_{\Upsilon^{\prime\prime}}/\beta_{\chi_{b0}^{\prime}}=0.776/0.945$, $0.876/0.945$, $0.976/0.945$, $1.076/0.945$, respectively; in (IV) $\chi_{b0}\to \Upsilon \gamma$, $\beta_{\Upsilon}=0.94\,\rm{GeV}$.}
\end{figure}

Finally, we present the results with two universal parameters $m_Q$ and $\beta_{Q\bar{Q}}$ (and one universal parameter  $\beta_{Q\bar{Q}}$) for completeness. If we take two universal parameters $m_c$ ($m_b$) and $\beta_{c\bar{c}}$ ($\beta_{b\bar{b}}$) to fit the experimental data on the charmonium (bottomnium) $E1$ transitions, the optimal values are
$m_c=1.5941\pm0.0073690\,\rm{GeV}$, $m_b=2.4125\pm 0.14607\,\rm{GeV}$,
 $\beta_{c\bar{c}}=1.2359\pm0.013862\,\rm{GeV}$,
  $\beta_{b\bar{b}}=0.58166\pm0.042596\,\rm{GeV}$, the resulting decay widths are
\begin{eqnarray}
 \Gamma_{\chi_{c0}\to J/\psi \gamma}&=&122.87^{+14.62}_{-13.86}{}^{+3.93}_{-3.76}\,\rm{KeV} \, , \nonumber \\
\Gamma_{\psi^{\prime}\to \chi_{c0} \gamma}&=&27.68^{+0.52}_{-0.52} {}^{+2.03}_{-1.96} \,\rm{KeV} \, , \nonumber \\
\Gamma_{\Upsilon^{\prime}\to \chi_{b0} \gamma}&=&0.908^{+0.122}_{-0.105} {}^{+0.035}_{-0.050}\,\rm{KeV} \, ,\nonumber \\
\Gamma_{\Upsilon^{\prime\prime}\to \chi_{b0} \gamma}&=&0.064^{+0.025}_{-0.017} {}^{+0.025}_{-0.019}\,\rm{KeV} \, , \nonumber \\
\Gamma_{\Upsilon^{\prime\prime}\to \chi_{b0}^{\prime} \gamma}&=&1.355^{+0.100}_{-0.096} {}^{+0.033}_{-0.066}\,\rm{KeV}  \, ,
\end{eqnarray}
where the uncertainties come from the heavy quark masses and shape parameters, sequentially.
The numerical values of the decay widths  are compatible with the experimental data within uncertainties, the  $m_c$ reaches  upper bound of the usually used  constituent quark mass $m_c=(1.4-1.6)\,\rm{GeV}$ with $2m_c> M_{J/\psi}$, while the $m_b$ is about one-half of the usually used  constituent quark mass, and it  is unacceptable. Furthermore, tiny  uncertainty $\delta m_c=\pm0.0073690\,\rm{GeV}$ leads to rather larger uncertainty $\delta\Gamma_{\chi_{c0}\to J/\psi \gamma}={}^{+14.62}_{-13.86}\,\rm{KeV}$.  The parameters $m_c=1.5941\pm0.0073690\,\rm{GeV}$, $m_b=2.4125\pm 0.14607\,\rm{GeV}$ are not robust and discarded.

On the other hand, if we take the ideal heavy quark masses  $m_c=1.5 \,\rm{GeV}$ and $m_b=4.8\,\rm{GeV}$, and use one universal parameter  $\beta_{c\bar{c}}$ ($\beta_{b\bar{b}}$) to fit the experimental data on the charmonium (bottomnium) $E1$ transitions, the optimal values are
 $\beta_{c\bar{c}}=1.1490\pm0.011593\,\rm{GeV}$ and 
  $\beta_{b\bar{b}}=1.1418\pm0.044619\,\rm{GeV}$, the resulting decay widths are
\begin{eqnarray}
 \Gamma_{\chi_{c0}\to J/\psi \gamma}&=&344.25^{+2.08}_{-1.83} \,\rm{KeV} \, , \nonumber \\
\Gamma_{\psi^{\prime}\to \chi_{c0} \gamma}&=&21.96^{+1.56}_{-1.49} \,\rm{KeV} \, , \nonumber \\
\Gamma_{\Upsilon^{\prime}\to \chi_{b0} \gamma}&=&0.463^{+0.001}_{-0.005} \,\rm{KeV} \, ,\nonumber \\
\Gamma_{\Upsilon^{\prime\prime}\to \chi_{b0} \gamma}&=&0.057^{+0.010}_{-0.010} \,\rm{KeV} \, , \nonumber \\
\Gamma_{\Upsilon^{\prime\prime}\to \chi_{b0}^{\prime} \gamma}&=&0.545^{+0.024}_{-0.018} \,\rm{KeV}  \, ,
\end{eqnarray}
where the uncertainties come from the shape parameters. The  discrepancies between the theoretical and experimental values of the decay widths are huge for the radiative  transitions $\chi_{c0}\to J/\psi \gamma$, $\Upsilon^{\prime}\to \chi_{b0} \gamma$, $\Upsilon^{\prime\prime}\to \chi_{b0}^{\prime} \gamma$.
  The parameters  $\beta_{c\bar{c}}=1.1490\pm0.011593\,\rm{GeV}$ and
  $\beta_{b\bar{b}}=1.1418\pm0.044619\,\rm{GeV}$ are poor and discarded.

\section{Conclusions}

In  this article, we study the radiative transitions among the vector and scalar heavy  quarkonium states in the framework of the CLFQM.
In calculations, we observe that the radiative decay widths are sensitive to the constituent quark masses and the shape parameters of the LFWF.
We reproduce the experimental data for the observed processes   with suitable parameters, while the predictions for the un-observed processes are  consistent or
inconsistent with other theoretical calculations.
The parameters can be fitted to the precise experimental data in the  future.

\section*{Acknowledgment}
This  work is supported by National Natural Science Foundation of
China, Grant Number 11075053,  and the
Fundamental Research Funds for the Central Universities.

\end{document}